\documentclass[12pt,onecolumn,oneside,a4paper,peerreview]{IEEEtran}
\usepackage[T1]{fontenc}
\usepackage{graphicx,latexsym,subfigure,amsmath,epsfig,latexsym,subfigure,amsmath,amssymb,multirow,booktabs}
\usepackage{times}
\usepackage{cite}
\usepackage{fancyhdr}
\usepackage{algorithm,amsthm}
\usepackage{epstopdf}
\usepackage{cases}
\usepackage{color}
\usepackage{bm}
\usepackage{url}
\usepackage{multirow}
\usepackage{cases}
\usepackage{setspace}
\usepackage{algorithmicx}
\usepackage{textcomp}
\usepackage{amsfonts}
\usepackage{algpseudocode}
\begin{document}
\title{Hybrid Beamforming for mmWave MU-MISO Systems Exploiting Multi-agent Deep Reinforcement Learning}
\author{
\begin{minipage}{0.98\columnwidth}
\vspace*{0.4cm}
\begin{center}
\IEEEauthorblockN{Qisheng Wang\IEEEauthorrefmark{1},
              Xiao Li\IEEEauthorrefmark{1},
              Shi Jin\IEEEauthorrefmark{1},
              and Yijiain Chen\IEEEauthorrefmark{2}\\
\vspace*{0.2cm} \small{
\IEEEauthorblockA{\IEEEauthorrefmark{1}National Mobile Communications Research Laboratory,
Southeast University, Nanjing 210096, China\\
\IEEEauthorrefmark{2}ZTE Corporation, Shenzhen 518057, China} } }
\end{center}
\end{minipage}

\thanks{This manuscript has been accepted by IEEE Wireless Communications Letters.}
} \IEEEaftertitletext{\vspace{-0.75\baselineskip}}
\maketitle
\vspace*{0.8cm}


\begin{abstract}
In this letter, we investigate the hybrid beamforming based on deep reinforcement learning (DRL) for millimeter Wave (mmWave) multi-user (MU) multiple-input-single-output (MISO) system. A multi-agent DRL method is proposed to solve the exploration efficiency problem in DRL. In the proposed method, prioritized replay buffer and more informative reward are applied to accelerate the convergence. Simulation results show that the proposed architecture achieves higher spectral efficiency and less time consumption than the benchmarks, thus is more suitable for practical applications.
\end{abstract}
\vspace*{0.4cm}


%

\section{Introduction}
Hybrid beamforming (HBF) has been a key technique in the application of millimeter Wave (mmWave) massive multiple-input-multiple-output (MIMO) system to reduce the amount of radio-frequency (RF) chains. To obtain the hybrid precoding matrices, several iterative methods, such as \cite{el2014spatially,sohrabi2016hybrid,yu2016alternating,li2018joint}, have been proposed for single-user and multi-user (MU) systems.
While providing effective HBF solutions, these algorithms were based on the assumption that the array response sets of the transceiver are available. This requires the knowledge of the angles of arrival and departure (AoA/AoD) of each user, which are hard to accurately estimate in practice. Also, the iterative algorithms introduce unnegligible processing delays.

Recently, due to its ability to handle the non-convex problem, reinforcement learning (RL) has been used in wireless communication systems design \cite{feng2020DRL,lizarraga2019hybrid,peken2019reinforcement,9112250,nasir2019multi,de2018cooperative}. Compared to the supervised learning (SL) methods which are widely investigated these years, RL methods do not need the pre-obtained large amount of training data, which might be very difficult to obtain. Moreover, RL is more robust to the environment\cite{9112250}. For the SL methods, new training data is needed and the network needs to be retrained, when the transmission environment changes to the one not included in the training data. In contrast, RL can adaptively and efficiently track the environment change based on its experience buffer.
In \cite{lizarraga2019hybrid,peken2019reinforcement}, RL method was used to choose the HBF matrices from codebooks generated by traditional methods. In \cite{9112250}, single-agent deep RL (DRL) was used to design the digital precoder. Compared to single-agent DRL, multi-agent DRL (MADRL) algorithm can improve the learning speed and reduce the exploration cost. In \cite{nasir2019multi,de2018cooperative}, the Q-learning and deep Q-networks (DQN) were extended to multi-agent pattern to solve the power control and beamforming problems.

In this letter, we investigate the HBF design for mmWave MU-MISO system exploiting DRL method. We propose a deep deterministic policy gradient (DDPG)\cite{lillicrap2015continuous} based MADRL algorithm to learn the analog beamformers. The proposed algorithm employs multi-agent joint exploration, improved experience replay buffer with priority, and more informative reward to simultaneously explore different subspaces of the environment. Simulations show that the performance and convergence speed of the proposed MAHBF algorithm outperforms the traditional algorithms.

\section{System model}
Consider a mmWave MU-MISO system consisting of a base station (BS) with $N_t$ antennas and $N_{RF}^t$ RF chains, and $K$ single-antenna users, where $K\leq N_{RF}^t<N_t$. The 
received signal of user $k$ can be written as
\begin{equation}\label{eq12}
y_k=\mathbf{h}_k^H\mathbf{f}_{k}s_k+\mathbf{h}_k^H\sum\nolimits_{l\neq k}\mathbf{f}_{l}s_l+n_k,
\end{equation}
where $s_k$ satisfying $\mathbb{E}[s_k^2]=1$ and $n_k\sim\mathcal{CN}(0,\sigma_k^2)$ are the transmitted signal and received noise of user $k$, $\sigma_k^2$ is the noise power, $\mathbf{h}_k\in\mathbb{C}^{N_t\times 1}$ is the channel vector from BS to user $k$, $\mathbf{f}_k=\mathbf{F}_{RF}\mathbf{f}_{D_k}$, 
$\mathbf{f}_{D_k}\in\mathbb{C}^{N_{RF}^t\times 1}$ is the digital beamforming vector of user $k$, $\mathbf{F}_{RF}\in\mathbb{C}^{N_t\times N_{RF}^t}$ is the analog precoder with the $(i,j)$-th element $\mathbf{F}_{RF}(i,j)$ satisfying the constant modulus constraints $|\mathbf{F}_{RF}(i,j)|=1$, the beamforming vector satisfies the total power constraint $\mathrm{Tr}(\mathbf{F}^H_{RF}\mathbf{F}_{RF}\mathbf{F}_{D}\mathbf{F}^H_{D})\leq P_t$, $P_t$ is the transmitted power, and $\mathbf{F}_{D}=[\mathbf{f}_{D_1},\cdots,\mathbf{f}_{D_K}]$. Then, the spectral efficiency of user $k$ is
\begin{equation}\label{eqRk}
\begin{split}
 R_k=\log_{2}\left(1+\frac{|\mathbf{h}_k^H\mathbf{F}_{RF}\mathbf{f}_{D_k}|^2}{\sigma_k^2+\sum_{l\neq k}|\mathbf{h}_k^H\mathbf{F}_{RF}\mathbf{f}_{D_l}|^2}\right).
\end{split}
\end{equation}

Assuming uniform linear array (ULA) at the BS, we use the geometric channel model \cite{raghavan2010sublinear}, i.e.,
\begin{equation}\label{eqhk}
\mathbf{h}_k=\sqrt{\frac{N_t}{N_{cl}N_{ray}}}\sum_{i=1}^{N_{cl}}\sum_{j=1}^{N_{ray}}\alpha_{ij}\mathbf{g}_t(\varphi^t_{ij}),
\end{equation}
where $N_{cl}$ is the number of scattering clusters, $N_{ray}$ is the number of scattering rays per cluster, $\alpha_{ij}\sim\mathcal{CN}(0,\sigma^2_{\alpha,i})$ is the complex path gain of $j$-th ray in the $i$-th cluster, $\sigma^2_{\alpha,i}$ is the average power gain of the $i$-th cluster, $\varphi^t_{ij}$ is the AoD,
\begin{equation}\label{eqg}
\mathbf{g}_t(\varphi)=\frac{1}{N_t}[1,e^{j\frac{2\pi\bar{d}}{\lambda}sin(\varphi )},\cdots,e^{j(N_t-1)\frac{2\pi\bar{d}}{\lambda}sin(\varphi )}]^T,
\end{equation}
$\lambda$ is the carrier wavelength, and $\bar{d}$ is the antenna spacing.

To maximize the throughput of the considered system, the HBF design problem can be given as
\begin{equation}\label{eqmaxR}
\begin{aligned}
\mathop{\max}\limits_{\mathbf{F}_D,\mathbf{F}_{RF}}&\sum\nolimits_{k=1}^{K}R_k,\\
\mathrm{s.t.}~\mathrm{Tr}(\mathbf{F}^H_{RF}\mathbf{F}_{RF}&\mathbf{F}_{D}\mathbf{F}^H_{D})\leq P_t,~ |\mathbf{F}_{RF}(i,j)|=1, \forall i,j.
\end{aligned}
\end{equation}
In this letter, we try to solve the analog precoder design problem in (\ref{eqmaxR}) through MADRL algorithm, while adopt the zero-forcing (ZF) digital precoder to suppress the inter-user interference.

\section{MADRL Hybrid Beamforming Architecture}
In this section, we propose a MADRL algorithm to design the analog precoder for the considered MU-MISO systems.

\subsection{Overall Architecture}
\begin{figure}[htbp]
\centering\includegraphics[width=5.0in]{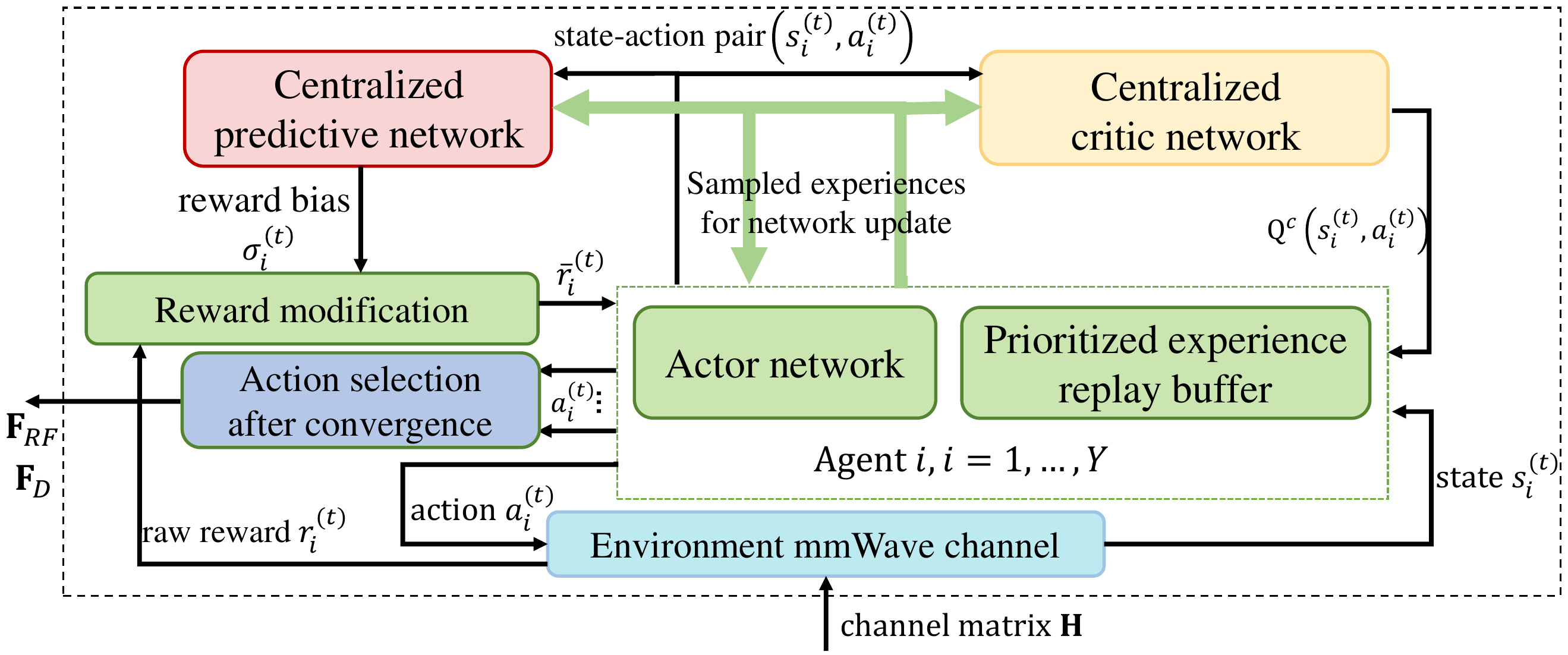}\\
\caption{Block diagram of the MADRL HBF method.}\label{fig5}
\end{figure}

The proposed MADRL HBF (MAHBF) algorithm regards the whole transmission system seen at BS as the environment. It takes the channel matrix $\mathbf{H}=[\mathbf{h}_1,\cdots,\mathbf{h}_K]^H$ as input and outputs the analog precoder and its corresponding digital precoder after several learning iterations. As illustrated in Fig. \ref{fig5}, the core of this algorithm consists of $Y$ agents, a centralized critic network to coordinate the behaviours of the agents, a centralized predictive network to guide the exploration of the agents. Each agent $i$ contains an actor network $\mathcal{A}_i$, a target actor network $\mathcal{A}'_i$, and a prioritized replay buffer $\mathcal{D}_i$ with capacity $N_{\mathcal{D}_i}$. The centralized critic network $\mathcal{C}$ and predictive network $\mathcal{P}$ also have a corresponding target network $\mathcal{C}'$ and $\mathcal{P}'$. The target networks are used to soft update these networks\cite{lillicrap2015continuous}. For simplicity, the target networks are not shown in Fig. \ref{fig5}. The netwrok $\mathcal{A}_i$, $\mathcal{A}_i'$, $\mathcal{C}$, $\mathcal{C}'$, $\mathcal{P}$, and $\mathcal{P}'$ are parameterized by $\bm{\theta}_{\mathcal{A}_i}$, $\bm{\theta}_{\mathcal{A}_i'}$, $\bm{\theta}_{\mathcal{C}}$, $\bm{\theta}_{\mathcal{C}'}$, $\bm{\theta}_{\mathcal{P}}$, and $\bm{\theta}_{\mathcal{P}'}$, respectively.

Note that our aim is to obtain the analog precoder, and each element of the analog precoder can be uniquely determined by its phase. Thus, at learning iteration $t$, the state of the $i$-th agent is defined as the phases of the elements of the analog precoder at the previous learning iteration, i.e., $s_{i}^{(t)}=\{\angle\mathbf{F}_{RF,i}^{(t-1)}\}$, and the input vector of its actor network is
\begin{equation}\label{eqs}
\begin{split}
\mathbf{s}^{(t)}_i=\mathrm{vec}(\angle\mathbf{F}_{RF,i}^{(t-1)}),
\end{split}
\end{equation}
where $\mathbf{F}^{(t-1)}_{RF,i}$ is the analog precoder obtained by the $i$-th agent at learning iteration $t-1$, and $\angle$ means the phases of the matrix elements. Its corresponding action is the phase of the analog precoder at the current learning iteration, i.e., $a_{i}^{(t)}=\{\angle\mathbf{F}_{RF,i}^{(t)}\}$. Thus, the output vector of the actor network is
\begin{equation}\label{eqa}
\begin{split}
\mathbf{a}^{(t)}_i=\mathrm{vec}(\angle\mathbf{F}_{RF,i}^{(t)}).
\end{split}
\end{equation}
The initial states $s_{i}^{(1)}$ of different agents are set to be orthogonal so that they are initialized at different subspaces. Specifically, all the $\mathbf{F}_{RF,i}^{(0)}$ are first randomly generated, and then orthogonalized by the Schmidt orthogonalization approach so that the initial actions of different agents satisfy that $(\mathrm{vec}(\mathbf{F}_{RF,i}^{(0)}))^H\mathrm{vec}(\mathbf{F}_{RF,j}^{(0)})=0,i\neq j$.

With the obtained $\mathbf{F}_{RF,i}^{(t)}$, ZF beamforming is used to generate the low-dimensional digital precoder $\mathbf{F}_{D,i}^{(t)}$\cite{sohrabi2016hybrid}, i.e.,
\begin{equation}\label{eqfzf}
\begin{split}
\mathbf{F}_{D,i}^{(t)} = \mathbf{\widetilde{F}}_{D,i}^{(t)} \big( \mathbf{P}^{(t)}_i \big)^{\frac{1}{2}},
\end{split}
\end{equation}
where $\mathbf{F}_{D,i}^{(t)}=[\mathbf{f}_{D_1,i}^{(t)},\cdots,\mathbf{f}_{D_K,i}^{(t)}]$, $\mathbf{f}_{D_k,i}^{(t)}$ is the digital beamforming vector of user $k$ obtained by the $i$-th agent,
\begin{equation}
\mathbf{\widetilde{F}}_{D,i}^{(t)} = \big(\mathbf{F}_{RF,i}^{(t)}\big)^{H} \mathbf{H}^H \big(\mathbf{H}\mathbf{F}_{RF,i}^{(t)} \big(\mathbf{H}\mathbf{F}_{RF,i}^{(t)} \big)^H \big)^{-1},
\end{equation}
$\mathbf{P}^{(t)}_i=\mathrm{diag}\big(p_{1,i}^{(t)},\cdots,p_{K,i}^{(t)}\big)$ with $p_{k,i}^{(t)}$ representing the received signal power obtained by user $k$ of the $i$-th agent. The diagonal power matrix $\mathbf{P}_i^{(t)}$ is obtained by the water-filling method \cite{sohrabi2016hybrid} to maximize the sum rate, and can be given as $p_{k,i}^{(t)}=({\mu_i^{(t)}}/{y_{k,i}^{(t)}}-\sigma_k^2)^+$, where $(\cdot)^+$ is the non-negative operator, $y_{k,i}^{(t)}$ is the $k$-th diagonal element of $\mathbf{Y}_i^{(t)}=(\mathbf{\widetilde{F}}_{D,i}^{(t)})^H(\mathbf{F}_{RF}^{(t)})^H\mathbf{F}_{RF,i}^{(t)}\mathbf{\widetilde{F}}_{D,i}^{(t)}$, and $\mu_i^{(t)}$ is selected to ensure the power constraint $\sum_ky_{k,i}^{(t)}p_{k,i}^{(t)}\leq P_t,\forall i$.

With the output action of the $i$-th agent, the environment feeds back a raw reward $r_i^{(t)}$ to evaluate the action. Since the object of this HBF design is to maximize the sum rate, the raw reward is therefore defined as the sum rate.
With ZF digital precoders and (\ref{eqRk}), the raw reward can be expressed as
\begin{equation}
r_i^{(t)}=\sum\nolimits_{k=1}^{K}\log_{2}(1+p_{k,i}^2/\sigma_k^2).
\end{equation}
The predictive network $\mathcal{P}$ takes the state-action pair $(\mathbf{s}_i^{(t)},\mathbf{a}_i^{(t)})$ as the input, and outputs a predicted reward value $\sigma_i^{(t)}$. The raw reward $r_i^{(t)}$ is then modified into $\bar{r}_i^{(t)}$ based on $\sigma_i^{(t)}$, which will be introduced later in Section \ref{sec:reward}. The centralized critic network $\mathcal{C}$ also takes $(\mathbf{s}_i^{(t)},\mathbf{a}_i^{(t)})$ as input, it outputs the Q-value of each agent's state-action pair, i.e., $Q^{\mathcal{C}}(\mathbf{s}_i^{(t)},\mathbf{a}_i^{(t)}),$ to the corresponding agent. Next, the $i$-th agent stores the experience $\{s_i^{(t)},a_i^{(t)},\bar{r}_i^{(t)},s_i^{(t+1)},\varphi_i^{(t)}\}$ into its replay buffer $\mathcal{D}_i$, where $\varphi_i^{(t)}$ is the priority of this experience and will be introduced in Section \ref{sec:buffer}. After that, $M_i$ samples are taken from $\mathcal{D}_i$, where $\sum M_i=M$ and $M$ is the total number of samples taken from all the agents' buffers, to update all the neutral networks based on the Q-values. Then, the algorithm moves to the next learning iteration until $|\mathbf{F}_{RF,i}^{(t)}-\mathbf{F}_{RF,i}^{(t-1)}|<\tau_{\rm{thres}}$ or $t=T$, where $\tau_{\rm{thres}}$ is a pre-defined threshold. At last, the action of the agent with the largest Q-value is selected as the analog precoder.
The main steps of the proposed algorithm are listed in Algorithm \ref{algorithm1}\footnote{All the steps in Algorithm \ref{algorithm1} are for one channel realization, and retraining is needed when the channel conditions change. However, for practical implementations, the BS does not need to wait until the algorithm reaches the optimal solution to serve the users. It can perform the training and serve the users with the output precoder of each learning iteration at the same time.}, and the key details are described in the following subsections.
\begin{algorithm}
\centering
\caption{MADRL-aided HBF algorithm}\label{algorithm1}
\begin{algorithmic}[1]
\State Input the channel matrix $\mathbf{H}$;\\
       Initialize $\mathcal{C}$, $\mathcal{P}$, and $\{\mathcal{A}_i,\mathcal{D}_i\}_{i=1}^Y$;\\
       Initialize all the state $s_i^{(1)}$ orthogonally;
\For {$t=1, 2, \cdots, T$}
\State Each agent outputs its action $a_i^{(t)}$;
\State Environment feedbacks reward $r_i^{(t)}$ to each agent;
\State Each agent outputs its $(s_i^{(t)},a_i^{(t)})$ pair to $\mathcal{C}$ and $\mathcal{P}$;
\State $\mathcal{C}$ outputs $Q^{\mathcal{C}}(s_i^{(t)},a_i^{(t)})$ to each agent;
\State $\mathcal{P}$ outputs $\sigma_i^{(t)}=\mathcal{P}(s_i^{(t)},a_i^{(t)})$ to each agent;
\State The reward is modified according to (\ref{eq46});
\State Each agent store experience in its replay buffer $\mathcal{D}_i$;
\State Sampling from the buffers to update $\mathcal{C}$, $\mathcal{P}$, and $\{\mathcal{A}_i\}$;
\State if $|\mathbf{F}_{RF,i}^{(t)}-\mathbf{F}_{RF,i}^{(t-1)}|<\tau_{\rm{thres}}$ then break;
\EndFor
\State Select the action $a_i^{(t)},\forall i$ with the largest Q-value as $\mathbf{F}_{RF}$ and the corresponding $\mathbf{F}_{D,i}^{(t)}$ as $\mathbf{F}_{D}$;
\end{algorithmic}
\end{algorithm}

\subsection{Multi-agent Joint Exploration}
Considering that a single agent can only explore a local subspace and needs high sampling complexity to learn an acceptable behaviour policy, one possible solution to accelerate the convergence of policy iteration is the joint exploration of the multiple agents. In the proposed MADRL algorithm, $Y$ agents are initialized orthogonally to hanlde the same task. Then, it coordinately explores the state space of each agent in the previous learning stage, so that each agent can explore different state space to speed up the convergence of the beamforming policy.

Note that the reinforcement learning works based on the assumption that the environment can be formulated as a Markov Decision Process (MDP). In the multi-agent learning system, each agent's state transition depends on the agents' joint actions, which means the environment of one agent may not be stationary as the other learning agents update their policies. Thus, the Markov property in the single-agent case no longer holds. To keep a stationary environment for all agents, there should be collaborative approach among the agents. Therefore, a centralized critic network $\mathcal{C}$ shared by all agents is introduced to ensure that the evaluation of different agents' actor networks are impartial, so that, from the perspective of a certain agent, the environment is stationary. Specifically, at every learning step, the critic network obtains the state-action pairs from all agents, outputs the Q-value to evaluate them based on the agents' observation, and feeds back the Q-values to each agent for their updates.

\subsection{Prioritized Replay Buffer}\label{sec:buffer}
The experience replay buffer enables the DRL agent to memorize and reuse prior experiences, and update its neural networks by uniformly sampling from the buffer. However, this method simply replays the samples at the same frequency, regardless of their significance. Since the networks are updated to minimize the temporal-difference error (TD-error) between the approximated Q-value $Q^{\mathcal{C}}(s_i,a_i)$ and the target $y_i$ by stochastic gradient descent (SGD), the transitions with larger TD-error will contribute more in calculating the gradient. Therefore, we demonstrate the importance of each transition by the TD-error \cite{Schaul2015Prioritized}, and intend to reuse the experiences with larger TD-error more frequently to make exploration efficient. Thus, the priority of the $n$-th transition in the replay buffer $\mathcal{D}_i$ is defined as the difference between its Q-value obtained from the critic network and the modified reward, i.e., $\varphi^{(t_n)}_i=Q^{\mathcal{C}}(s^{(t_n)}_i,a^{(t_n)}_i)-\bar{r}^{(t_n)}_i+\delta,$
and is also stored in the replay buffer, where the transition $e^{(n)}_i$$=$$\{s^{(t_n)}_i,a^{(t_n)}_i,\bar{r}^{(t_n)}_i,s^{(t_n+1)}_i,\varphi^{(t_n)}_i\}$ is the $n$-th experience in $\mathcal{D}_i$, which is the experience of the $t_n$-th learning iteration of the $i$-th agent, and $0<\delta\ll 1$ is a bias to ensure positive priority.

\begin{figure}[htbp]
\centering\includegraphics[width=5in]{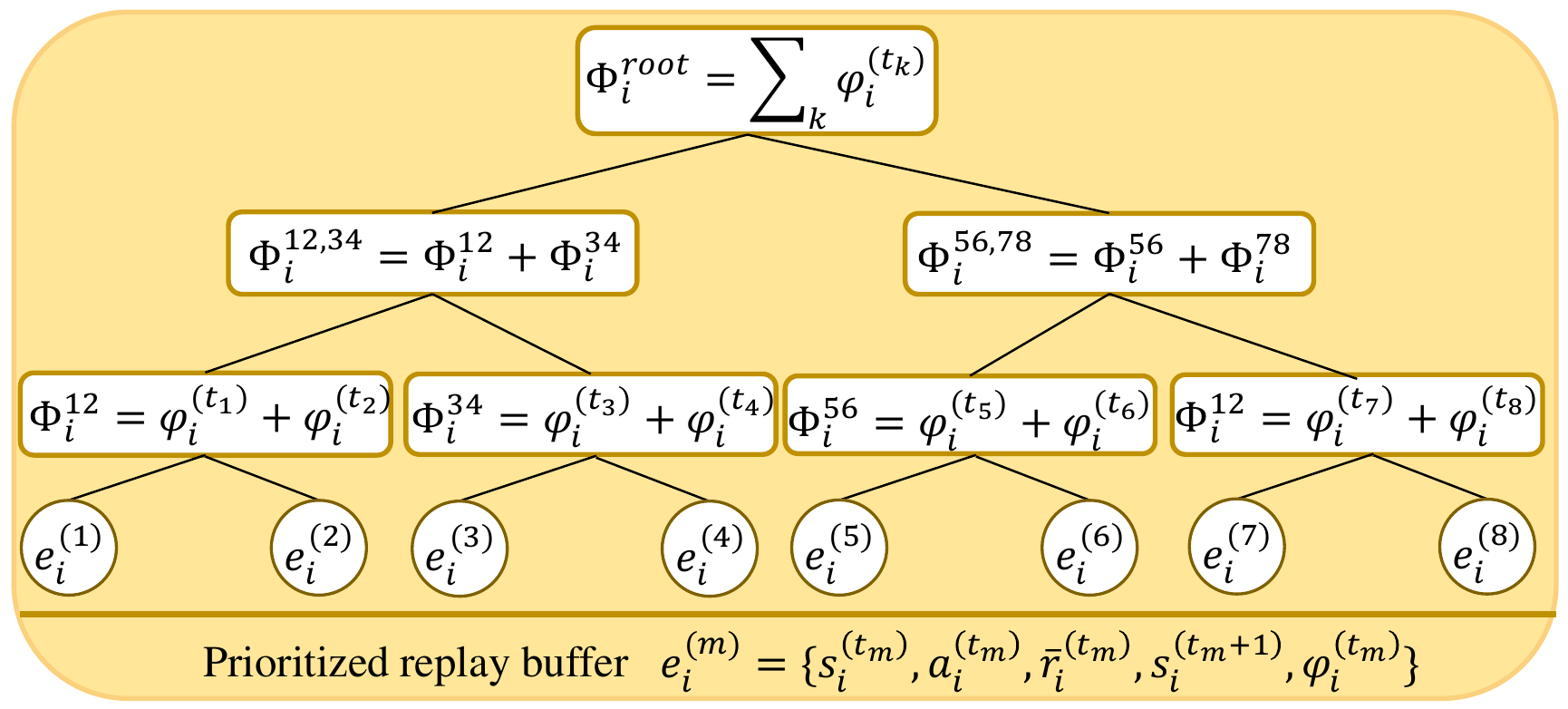}\\
\caption{The Prioritized Experience Replay}\label{fig6}
\end{figure}

With the transition $e_i^{(n)}$, each agent holds its own replay buffer $\mathcal{D}_i$ in the form of ``sum-tree'' to improve sampling efficiency, as shown in Fig. \ref{fig6}. The lowest-level leaf node stores the transition while the remaining nodes only store the sum of the priority of their children nodes, i.e., $\Phi_i^{ab}$, where $a$ and $b$ are the indexes of its children nodes. The root node records the sum of the priority of all samples in $\mathcal{D}_i$, denoted as $\Phi_i^{root}$. Considering the access frequency of a sample can also reflect its importance \cite{dai2019reinforcement}, we further modify the priority of each leaf node using its access frequency $\rho^{(n)}_i$ as
\begin{equation}\label{eq39}
\varphi^{(t_n)}_i=Q^{\mathcal{C}}(s^{(t_n)}_i,a^{(t_n)}_i)-r^{(t_n)}_i+\rho^{(n)}_i/\sum\nolimits_{j}\rho^{(j)}_i+\delta.
\end{equation}

To make each agent effectively utilize the knowledge learned by the other agents instead of exploring the entire state-action space, we further take the contribution of different agent into consideration. Note that the sum priority $\Phi_i^{root}$ stored in the root node of $\mathcal{D}_i$ reveals the importance of the $i$-th agent in the update of the centralized critic network. The agent with higher $\Phi_i^{root}$ should contribute more. Thus, we define the priority of the $i$-th agent as
$q_i=\exp(\Phi_i^{root})/\sum_{j}\exp(\Phi_j^{root})$.

In each learning iteration, a $M$-size minibatch is sampled, and the number of experiences sampled from $\mathcal{D}_i$ is $M_i=\left\lfloor q_iM\right\rfloor$, where $\left\lfloor \cdot\right\rfloor$ is the round down operation. Moreover, the probability of sampling the $m$-th transition in the $i$-th replay buffer $\mathcal{D}_i$ is set to $P_i^{(m)}=\varphi_i^{(t_m)}/\sum_{i,j}\varphi_i^{(t_j)}$, so that the probability of being sampled is monotonic in the transition's priority. The loss function of the centralized critic network is
\begin{equation}\label{eq42}
L(\bm{\theta}_{\mathcal{C}})=\frac{1}{M}\sum_{i=1}^{Y}\sum_{m=1}^{M_i}q_i^{(t)}\Big(Q^{\mathcal{C}}(s_i^{(t_m)},a_i^{(t_m)})-y_i^{(t_m)}\Big)^{2},
\end{equation}
where $q_i^{(t)}$ is the priority of the $i$-th agent at learning iteration $t$, and the target
\begin{equation}\label{eq43}
y_i^{(t_m)}=r_i^{(t_m)}+\gamma Q^{\mathcal{C}'}\Big(s_i^{(t_m+1)},a_i^{(t_m')}\Big)|_{a_i^{(t_m')}=\mathcal{A}_i'(s_i^{(t_m+1)})},
\end{equation}
where $\gamma$ is the discount factor to guarantee convergence. The policy gradient of the $i$-th agent and the centralized critic network are updated via (\ref{eq42}) and
\begin{equation}\label{eq44}
  \bm{\theta}_{\mathcal{A}_i}=\mathop{\arg\min}_{\bm{\theta}_{\mathcal{A}_i}}\frac{q_i^{(t)}}{M_i}\sum_{m=1}^{M_i}-Q^{{\mathcal{C}}}(s_i^{(t_m)},a)|_{a=\mathcal{A}_i'(s_i^{(t_m)})},
\end{equation}
and all the target networks are soft updated according to
\begin{equation}\label{eq_soft_update}
\begin{aligned}
\bm{\theta}_{\mathcal{A}_i'}=\tau\bm{\theta}_{\mathcal{A}_i}+(1-\tau)\bm{\theta}_{\mathcal{A}_i'},\bm{\theta}_{\mathcal{C}'}=\tau\bm{\theta}_{\mathcal{C}}+(1-\tau)\bm{\theta}_{\mathcal{C}'},
\end{aligned}
\end{equation}
where $\tau\ll1$ is an update factor to ensure the weights of target networks to change slowly. In this way, the agents with larger priority and the experiences with greater TD-errors are used to provide more information for the update of networks. The improved experience replay buffer enables the critic network to evaluate and coordinate all the agents' behavior, while the actor networks of different agents can learn concurrently. Thus, the time consumption of convergence is shortened.

\subsection{More informative reward}\label{sec:reward}
An undeniable problem in RL is the insufficient-information reward compared with the informative label in supervised learning. When most agents get feedback with insufficient information, the learning process is difficult to perform, especially under the unstationary environment caused by the interaction of multiple agents. A method for increasing the information embedded in the reward based on the latent state was proposed in \cite{vezzani2019learning}. It uses the experience stored from previous trajectories as the representation of reward to train a network, which predicts the reward of the new state-action pair. However, it requires numerous repetitive simulations, which is too costly in complex environments.

In this letter, we propose a centralized predictive network $\mathcal{P}$. It uses the output of the critic network as target to estimate the reward of the current state-action pair, therefore can gather the experiences from all agents' replay buffers for real-time training other than the pre-experiment in \cite{vezzani2019learning}. At every learning iteration $t$, the $i$-th agent obtains the state and action of the current learning iteration, i,e., $(s_{i}^{(t)},a_{i}^{(t)})$. The predictive network uses it as input, and outputs a predicted value $\sigma_i^{(t)}=\mathcal{P}(s_i^{(t)},a_i^{(t)}|\bm{\theta}_{\mathcal{P}})$ as bias to refine the reward, i.e.,
\begin{equation}\label{eq46}
\bar{r}_i^{(t)}=r_i^{(t)}+\eta\sigma_i^{(t)},
\end{equation}
where the discount factor $\eta$ is used to determine how much the predictive value is used to increase the information in the reward. The loss function of the predictive network is
\begin{equation}\label{eq47}
L(\bm{\theta}_{\mathcal{P}})=\frac{1}{M}\sum_{i=1}^{Y}\sum_{m=1}^{M_i}q_i^{(t)}(Q^{\mathcal{C}}(s_i^{(t_m)},a_i^{(t_m)})-\sigma_i^{(t_m)})^{2}.
\end{equation}
Then, it is updated by SGD. Note that the predictive and critic network are updated at the same time using the same experiences, and the gradient flows from the predictive network and actor networks to the critic network. Therefore, the final gradient of the critic network is the sum of the gradients from the predictive network and actor networks. In this way, the more informative reward accelerates the update of the centralized predictive network and critic network, which leads to more efficient and effective explorations of all agents.

\section{Simulation}
In this section, we provide the simulation results to validate the performance of the proposed MAHBF algorithm. The proposed algorithm and the two benchmarks are all implemented on the following configurations: deep learning platform PyTroch 1.3.1, one graphic for NVIDIA Tesla V100 32GB. In the simulations, we set $N_t=64$, $N_{RF}^t=K=8$, $\bar{d}=\lambda/2$, $N_{cl}=10$, and $N_{ray}=8$. All the networks are composed of four fully-connected layers. The input layer of the actor networks includes $N_tN_{RF}^t$ nodes, and the input layer of the critic and predictive networks both includes $2N_tN_{RF}^t$ nodes. The second and third layers of all the networks are hidden layers with 300 and 200 neurons, respectively. All the first three layers use the rectified linear units (ReLU) as the activation function. The output layer of each actor network has $N_tN_{RF}^t$ nodes, and the output layer of the critic and predictive network both have 1 node. All these output layers use tahn function as the activation function, and $\alpha=10^{-3}$, $\gamma=0.95$, $\tau=10^{-3}$, $\tau_{\rm{thres}}=10^{-4}$, $N_{\mathcal{D}_i}=500, \forall i$, and $M=32$.

\begin{figure}[htbp]
\centering\includegraphics[width=5.0in]{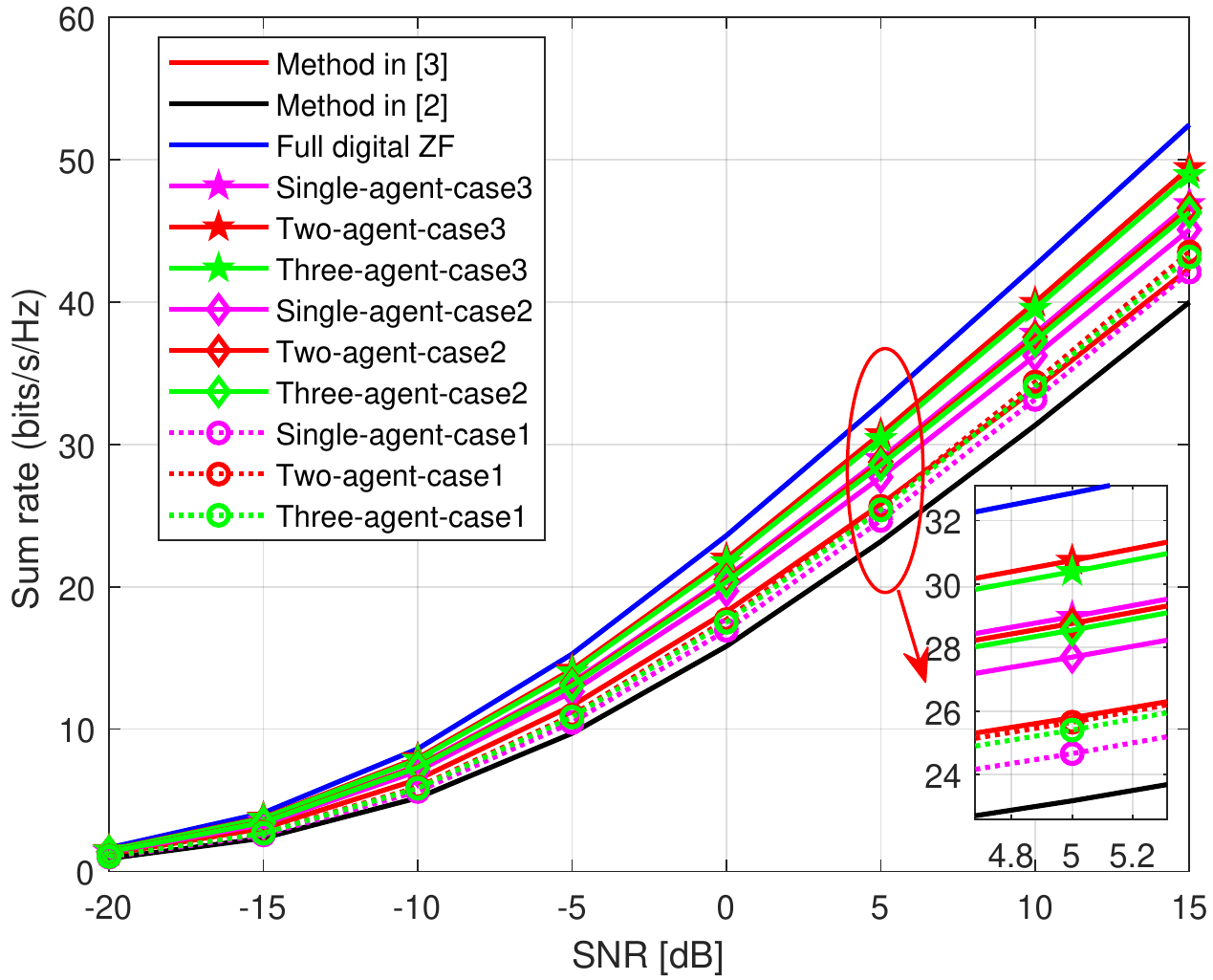}\\
\caption{Sum rate comparison of different precoding algorithms.}\label{fig10}
\end{figure}

\begin{figure}[htbp]
\centering\includegraphics[width=5.0in]{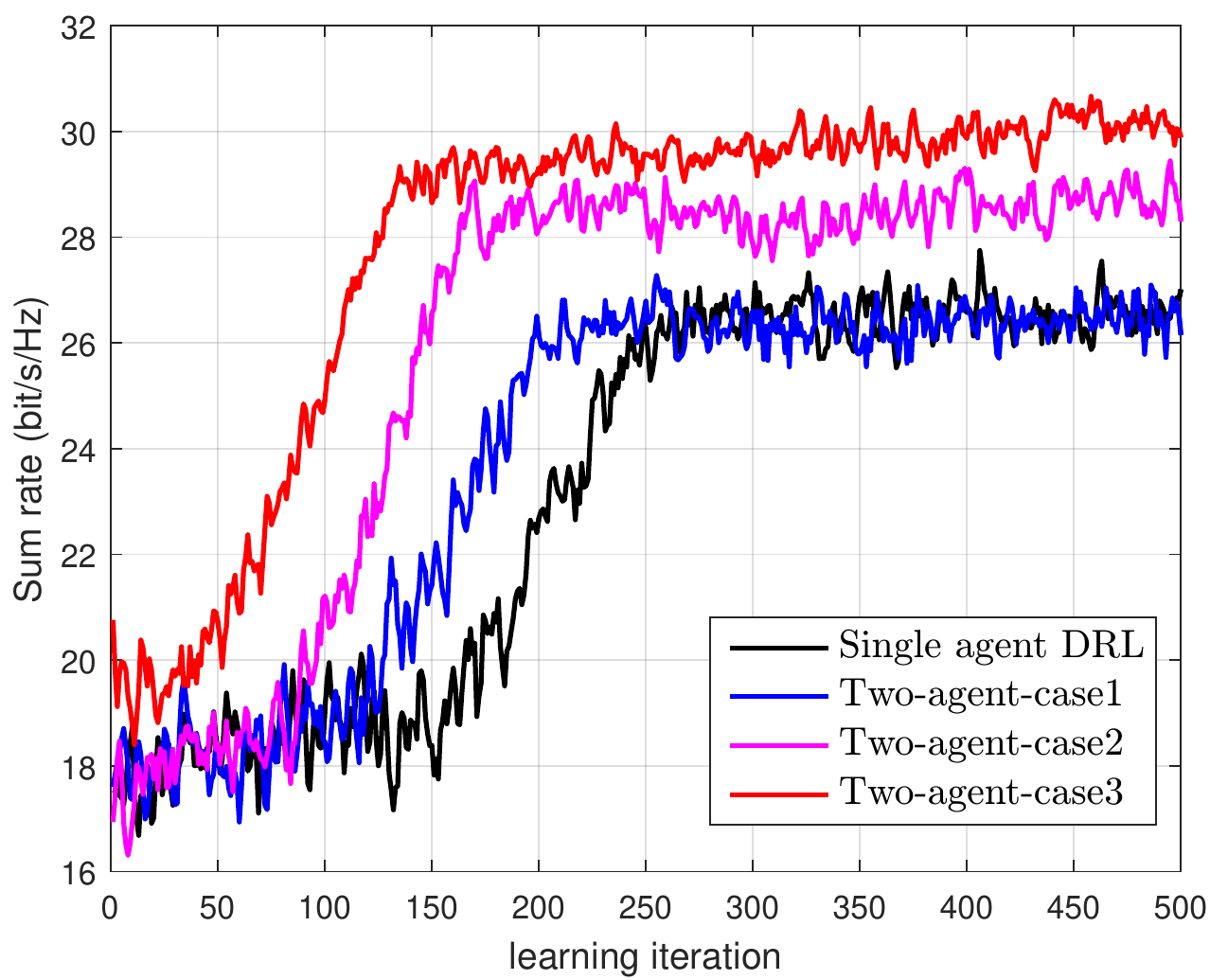}\\
\caption{Convergence performance of MAHBF with different improvements.}\label{fig11}
\end{figure}

Fig. \ref{fig10} compares the sum rate achieved by the proposed MAHBF algorithm under $Y=1,2$, and $3$, with the algorithms in \cite{sohrabi2016hybrid} and \cite{yu2016alternating}, and the performance of the full digital ZF precoder is provided as an upper bound. In this figure, the effects of each of the three proposed improvements are also shown. The algorithm with only the multi-agent exploration is referred to as ``case1'', the one with the multi-agent exploration and the prioritized replay buffer is referred to as ``case2'', and the one with all the three improvements is referred to as ``case3''. Fig. \ref{fig11} compares the convergence performance of these cases with the single-agent DRL algorithm, that does not apply these improvements, under $\mathrm{SNR} = 5~\mathrm{dB}$.

Fig. \ref{fig10} shows that almost all these three cases can achieve higher sum rate than the methods in \cite{sohrabi2016hybrid,yu2016alternating}, with the only exception that case1 is slightly inferior to the method in \cite{sohrabi2016hybrid} under low SNR with $Y=1$. Fig. \ref{fig11} shows that when $Y=2$, the convergence of case1, case2, and case3 requires about 200, 150, and 130 learning iterations, while the single-agent approach requires 250 learning iterations. From both figures, it can be seen that all these proposed improvements contribute to the sum rate increment as well as convergence acceleration, i.e., case3 outperforms case2, and case2 outperforms case1. It can also be seen that the prioritized buffer provides more performance increment than the other two improvements.

From Fig. \ref{fig10}, it can be seen that the performance of the proposed algorithm with all three improvements is very close to the upper bound. It can also be seen that the proposed algorithm with $Y=3$ outperforms the proposed algorithm with $Y=1$, while the proposed algorithm with $Y=2$ outperforms both the other two cases. This indicates that more agents is not always better, since the variance of the policy gradient can be increasingly large when the number of DRL agents increases\cite{lowe2017multi}. Although the proposed multi-agent exploration and the other two methods help to improve the system performance, this increasing variance of the policy gradient degrades the performance. Thus, there exists an optimal number of agents.

Table \ref{table2} shows the time consumption required for convergence of different algorithms when $\mathrm{SNR} = 5~\mathrm{dB}$. It can be seen that the time consumption of the proposed algorithm is much less than the other algorithms\footnote{Although the convergence time of the proposed algorithm might not be extremely short to satisfy the coherence time constraint of mmWave channel under current condition, we believe that it can be greatly reduced with the improving of the computing power brought by the development of the high performance computer.}. We also obtained that the convergence of two-agent and three-agent approaches requires about 130 and 145 learning iterations, while the single-agent approach requires 250 learning iterations. It can be seen that with multiple agents, the time consumption and iterations required to converge are reduced, and the proposed algorithm with $Y=2$ has the lowest time consumption.
\begin{table}[]
\centering
\setlength{\belowcaptionskip}{5pt}\caption{Time consumption (ms) of different algorithms}\label{table2}
\begin{tabular}{|c|c|c|c|c|}
\hline
Y=1 & Y=2  & Y=3   & Method in [3] & Method in [2] \\ \hline
159 & 85.7 & 109.3 & 792           & 61012         \\ \hline
\end{tabular}
\end{table}

\section{Conclusion}
In this letter, we proposed a novel hybrid beamforming design architecture based on MADRL algorithm for mmWave MU-MISO systems. In this algorithm, multiple agents were used to accelerate the learning process. Moreover, multi-agent joint exploration, an improved prioritized replay buffer, and a reward-prediction network were proposed. Simulations verified that the proposed algorithm can achieve considerable performance while has much less time consumption.


\bibliography{reference}

\begin{thebibliography}{10}

\bibitem{el2014spatially}
O.~El~Ayach, S.~Rajagopal, S.~Abu-Surra, Z.~Pi, and R.~W. Heath, ``Spatially
  sparse precoding in millimeter wave {MIMO} systems,'' {\em IEEE Trans.
  Wireless Commun.}, vol.~13, no.~3, pp.~1499--1513, 2014.

\bibitem{sohrabi2016hybrid}
F.~Sohrabi and W.~Yu, ``Hybrid digital and analog beamforming design for
  large-scale antenna arrays,'' {\em IEEE J. Sel. Top. Sign. Proces.}, vol.~10,
  no.~3, pp.~501--513, 2016.

\bibitem{yu2016alternating}
X.~Yu, J.-C. Shen, J.~Zhang, and K.~B. Letaief, ``Alternating minimization
  algorithms for hybrid precoding in millimeter wave {MIMO} systems,'' {\em
  IEEE J. Sel. Top. Sign. Proces.}, vol.~10, no.~3, pp.~485--500, 2016.

\bibitem{li2018joint}
Z.~Li, S.~Han, S.~Sangodoyin, R.~Wang, and A.~F. Molisch, ``Joint optimization
  of hybrid beamforming for multi-user massive {MIMO} downlink,'' {\em IEEE
  Trans. Wireless Commun.}, vol.~17, no.~6, pp.~3600--3614, 2018.

\bibitem{feng2020DRL}
K.~Feng, Q.~Wang, X.~Li, and C.-K. Wen, ``Deep reinforcement learning based
  intelligent reflecting surface optimization for {MISO} communication
  systems,'' {\em IEEE Wireless Commun. Lett.}, vol.~9, no.~5, pp.~745--749,
  2020.

\bibitem{lizarraga2019hybrid}
E.~M. Lizarraga, G.~N. Maggio, and A.~A. Dowhuszko, ``Hybrid beamforming
  algorithm using reinforcement learning for millimeter wave wireless
  systems,'' in {\em Proc. of RPIC}, pp.~253--258, 2019.

\bibitem{peken2019reinforcement}
T.~Peken, R.~Tandon, and T.~Bose, ``Reinforcement learning for hybrid
  beamforming in millimeter wave systems,'' International Foundation for
  Telemetering, 2019.

\bibitem{9112250}
Q.~{Wang}, K.~{Feng}, X.~{Li}, and S.~{Jin}, ``Precodernet: Hybrid beamforming
  for millimeter wave systems with deep reinforcement learning,'' {\em IEEE
  Wireless Commun. Lett.}, vol.~9, no.~10, pp.~1677--1681, 2020.

\bibitem{nasir2019multi}
Y.~S. Nasir and D.~Guo, ``Multi-agent deep reinforcement learning for dynamic
  power allocation in wireless networks,'' {\em IEEE J. Sel. A. Commun.},
  vol.~37, no.~10, pp.~2239--2250, 2019.

\bibitem{de2018cooperative}
C.~de~Vrieze, S.~Barratt, D.~Tsai, and A.~Sahai, ``Cooperative multi-agent
  reinforcement learning for low-level wireless communication,'' {\em arXiv
  preprint arXiv:1801.04541}, 2018.

\bibitem{lillicrap2015continuous}
T.~P. Lillicrap, J.~J. Hunt, A.~Pritzel, N.~Heess, T.~Erez, Y.~Tassa,
  D.~Silver, and D.~Wierstra, ``Continuous control with deep reinforcement
  learning,'' {\em arXiv preprint arXiv:1509.02971}, 2015.

\bibitem{raghavan2010sublinear}
V.~Raghavan and A.~M. Sayeed, ``Sublinear capacity scaling laws for sparse
  {MIMO} channels,'' {\em IEEE Trans. Inform. Theory}, vol.~57, no.~1,
  pp.~345--364, 2010.

\bibitem{Schaul2015Prioritized}
T.~Schaul, J.~Quan, I.~Antonoglou, and D.~Silver, ``Prioritized {E}xperience
  {R}eplay,'' {\em arXiv preprint arXiv:1511.05952}, 2015.

\bibitem{dai2019reinforcement}
C.~Dai, L.~Xiao, X.~Wan, and Y.~Chen, ``Reinforcement {L}earning with {S}afe
  {E}xploration for {N}etwork {S}ecurity,'' in {\em IEEE Proc. of ICASSP},
  pp.~3057--3061, 2019.

\bibitem{vezzani2019learning}
G.~Vezzani, L.~Gupta, and P.~Abbeel, ``Learning latent state representation for
  speeding up exploration,'' {\em arXiv preprint arXiv:1905.12621}, 2019.

\bibitem{lowe2017multi}
R.~Lowe, Y.~I. Wu, A.~Tamar, J.~Harb, O.~P. Abbeel, and I.~Mordatch,
  ``Multi-agent actor-critic for mixed cooperative-competitive environments,''
  in {\em Proc. of NeurIPS}, pp.~6379--6390, 2017.

\end{thebibliography}
\bibliographystyle{ieeetr}
\end{document}